\date{}
\documentclass[aps,pra,showpacs]{revtex4}
\usepackage{graphicx,psfrag,amsmath,amssymb,amsfonts,latexsym,color,dcolumn}
\begin{document}
\title{Quantum dissipative effects in moving imperfect mirrors: sidewise and normal motions}
\author{C\'esar D. Fosco$^{1,2}$ \footnote{fosco@cab.cnea.gov.ar}}
\author{Fernando C. Lombardo$^3$ \footnote{lombardo@df.uba.ar}}
\author{Francisco D. Mazzitelli$^{1,3}$ \footnote{fmazzi@df.uba.ar}}

\affiliation{$^1$ Centro At\'omico Bariloche
Comisi\'on Nacional de Energ\'\i a At\'omica, 
R8402AGP Bariloche, Argentina}
\affiliation{$^2$ Instituto Balseiro, Comisi\'on Nacional de Energ\'\i a At\'omica, 
R8402AGP Bariloche, Argentina}
\affiliation{$^3$ Departamento de F\'\i sica {\it Juan Jos\'e
 Giambiagi}, FCEyN UBA, Facultad de Ciencias Exactas y Naturales,
 Ciudad Universitaria, Pabell\' on I, 1428 Buenos Aires, Argentina }

\date{today}

\begin{abstract}  
We extend our previous work on the functional approach to the dynamical
Casimir effect, to compute dissipative effects due to the relative motion
of two flat, parallel, imperfect mirrors in vacuum.  
The interaction between the internal degrees of freedom of the mirrors and 
the vacuum field is modeled with a 
nonlocal term in the vacuum field action.
We consider two
different situations: either the motion is  `normal', i.e., the mirrors advance or recede
changing  the
distance $a(t)$ between them; or it is `parallel', namely,
$a$ remains constant, but there is a relative sliding motion of the
mirrors' planes.  For the latter, we show explicitly that there is a
non-vanishing frictional force, even for a constant shifting speed.
\end{abstract}

\maketitle
\section{Introduction}\label{sec:intro} 
Interesting manifestations of the vacuum electromagnetic field fluctuations
may arise when a neutral body (`mirror') is subjected to the influence of
certain time-dependent external conditions. A nice particular example of
this kind of phenomenon occurs when those varying conditions amount to a
{\em motion\/} of the body. When this body is an accelerated mirror, this
is the celebrated dynamical Casimir effect (DCE), or `motion induced
radiation', whereby  real photons are created out of the vacuum.  As with
any radiation phenomenon, it can be described from at least two points of
view: for the external agent driving the body, this process is perceived as
the cause of a dissipative force (which reacts against the change in the
external conditions), while, on the other hand, an observer measuring
electromagnetic field properties detects creation of real photons out of
the vacuum. 

Under the currently accessible experimental conditions, both the
dissipative force and the number of photons that may be created by a single
accelerated mirror are  deceptively small.  The main reason is that (for an
oscillatory motion) it would be necessary to attain prohibitively high
mechanical frequencies for the effect to be detected. 
There are, however, some experimental setups where the effect may be
purposely enhanced; for example, when a moving mirror is part of an
electromagnetic cavity, the mechanical oscillations of the mirror may
resonate with the corresponding normal cavity modes. In that case,
parametric amplification produces an exponential growth in the number
of emitted photons, at least for the idealized case of `perfect' mirrors,
namely, ones that behave as ideal conductors. 
For this kind of configuration, the estimated effect is much closer to
experimental verification and there are, indeed, several ongoing
experiments with that goal in mind. 

On the other hand, some alternative proposals invoke the use of
time-dependent changes in the electromagnetic {\em properties\/} of the
body, whereby mechanical motion is altogether avoided.  For general reviews
of these, and other aspects of the dynamical Casimir effect, see, for
example~\cite{reviews}.

An interesting particular case of motion induced effects is the so-called
`quantum friction', also due to the vacuum electromagnetic fluctuations,
where the theory predicts the appearance of non-contact frictional forces
between neutral bodies in relative sidewise motion.  This effect can be
understood in terms of the interchange of virtual photons between the
bodies, that then produce excitations of their internal degrees of freedom.
This quantum friction has been analyzed~\cite{Pendry97} (and
debated~\cite{debate}) at length, mainly  for the case of half-spaces
shifting with constant velocity (see also Refs. in \cite{others}, where two 
atoms on parallel trajectories, or an atom moving on a half-space are 
considered as a first approach to the more general situation of emission 
of light from sheared dielectric surfaces).

Therefore, dissipative effects on  moving bodies may not only be produced
by the excitation of real photons out of the quantum vacuum, but also of
the mirror's internal degrees of freedom, by the  mediation of the vacuum
field (in this case by virtual photons).  Note, however, that there is
another source of quantum dissipation, independent of the coupling to the
electromagnetic field, which is the excitation of the internal degrees of
freedom due to `fictitious', inertial forces~\cite{inertial} in an
accelerated system.

In a previous paper~\cite{Fosco:2007nz} we used a functional approach to
study the DCE for a single zero-width mirror, in a model where the
interaction between the internal degrees of freedom and the vacuum field
was described by a local $\delta$-potential term in the vacuum field
action, which had support on the (time-dependent) mirror's position.
Although sufficient to explain the production of real photons out of the
vacuum, this model has no room to describe processes where the mirror's
degrees of freedom are excited.  In that respect, one must consider more
general models. For example, for zero-width mirrors, the model should
include a {\em nonlocal\/} term quadratic in the vacuum field. Indeed, this
is the kind of term one naturally derives by integrating out microscopic
degrees of freedom in some models~\cite{plb08}. But one could also argue,
on more general grounds, that a (space or time) nonlocality must exist (in
a zero-width term) since most microscopic models shall indeed produce a
nontrivial momentum dependence in the interaction term, which is seen as a
certain nonlocality in spacetime.  

With the purpose to overcome that limitation in mind, in this paper we
analyze dissipative effects on moving mirrors, using a functional
formalism, generalizing our previous results ~\cite{Fosco:2007nz} in
several directions. First, we consider nonlocal boundary interaction
terms in the vacuum field action, to allow for a better description of the
effects of the microscopic degrees of freedom. We do that for either one or
two flat mirrors, both for parallel and normal motion.  This will allow us
to describe, on the same footing, dissipative forces coming from vacuum
field or matter (microscopic) degrees of freedom.  In particular, we shall
show that non-contact frictional forces do appear between thin mirrors,
even for parallel motion at a constant speed. Then, quantum friction induced 
by the sideway motion between dielectric mirrors is obtained as a byproduct of 
the effective action for the mirror coordinates. This effective action is the 
result of tracing out quantum fluctuations of the vacuum field and the 
internal degrees of freedom in the plates.

The structure of this paper is as follows: in~\ref{sec:model} we introduce
the model which is used in the rest of the paper as a convenient prototype
to study the effects mentioned above: it consists of  a quantum scalar
field coupled to the microscopic degrees of freedom of one or two imperfect
flat mirrors. The corresponding effective action due to the mirrors can be
formally written in terms of a functional  determinant, which in turn
depends on the scalar field (via a function of its free propagator) and
also on kernels that encode the properties of the mirrors.

In Section~\ref{sec:normal}, the general formal expression for the effective
action is evaluated  perturbatively for two mirrors in a normal motion
situation.
We present some explicit calculations in $1+1$ dimensions for
perfect and imperfect mirrors,
 showing that the imaginary part of the effective action peaks when the motion of the mirrors
is in resonance with the standing waves of the cavity.  
In Section~\ref{sec:friction}, we compute the effective action for
parallel motion. We first show that, for the interaction term to describe a
zero width mirror where the vacuum field propagates at a different speed
than in the vacuum, such interaction term must necessarily be nonlocal.
Then we show that, in that situation, nontrivial dissipative effects
appear, even for parallel motion with constant velocity. We also
present the perturbative form of the effective action for non-constant parallel
velocity of one of the mirrors.
Section~\ref{sec:concl} contains our conclusions.
\section{Imperfect zero-width mirrors coupled to a real scalar
field}\label{sec:model}
\subsection{One mirror}\label{ssec:onemirror}
Let us first consider a single, flat, non-relativistic mirror, coupled
to a real scalar field $\varphi$, in $3+1$ dimensions.  
We shall setup our conventions for Euclidean (imaginary time) spacetime,
although we shall occasionally undo the Wick's rotation to calculate the imaginary part of the real time effective
action.  Coordinates shall be denoted by  $x_\mu$ ($\mu=0,1,2,3$), and
the reference system is chosen in such a way that, at any given time, the
mirror occupies an $x_3\,=\,{\rm constant}$ plane, where the constant
will generally be time-dependent. Besides, the coordinates on those planes
are denoted by $x_\parallel \equiv (x_1,x_2)$.  Note that, under purely parallel
motion, the plane is invariant, but the coordinates $x_\parallel$ of the
points of the mirror will change. We shall only consider cases where the
sliding motion has a constant direction, thus, one can add to the choice of
reference system the extra requirement that only the $x_1$ coordinate of
points on the mirror can carry a time dependence.
Thus, (purely) normal motion of the mirror will be described by a single
function of time, \mbox{$x_3=q_\perp (x_0)$}, say; a single function is also
sufficient for purely parallel motion, we shall denote it
by~\mbox{$x_1=q_\parallel (x_0)$}.

The vacuum field $\varphi$ is coupled to the internal degrees of freedom of
the mirror, that we will denote 
generically by $\psi$, living on the
$2+1$ dimensional spacetime swept by the mirror in its time evolution.  The effective
action $\Gamma$ for the mirror coordinates, due to the quantum fluctuations of both
$\varphi$ and $\psi$, will be a functional of $q_\perp$ and
$q_{\parallel}$. It may be explicitly written, in terms of an Euclidean
functional integral:
\begin{equation} 
e^{-\Gamma(q_{\perp},q_\parallel)}=\int{\mathcal
D}\varphi{\mathcal D}\psi\,
e^{-S_0(\varphi)-S^{(0)}_m(\psi)-S^{(\rm int)}_m(\varphi,\psi)}\; , 
\end{equation}
where $S_0$ is the action of the vacuum field in the absence of
the mirror; which corresponds to a free real scalar field:  
\begin{equation}
S_0\;=\; \frac{1}{2} \int d^4x \,\big[ (\partial\varphi)^2 \,+\, m^2
\varphi^2 \big] \;.
\end{equation}
On the other hand, $S^{(0)}_m$ denotes the free action for the microscopic
degrees of freedom, while $S^{(\rm int)}_m$ is the term that couples them to
the vacuum field.

A convenient approach to find an expression for the effective action
$\Gamma$ is to perform the integration of the two relevant fields in two
successive steps. Considering first the microscopic degrees of freedom on the
mirrors, and under the usual assumption that the medium behaves linearly, one
keeps just the quadratic term in the result of that integral,
\begin{equation}\label{eq:pint}
 e^{-\Gamma(q_{\perp},q_\parallel)}=\int{\mathcal D}\varphi\,  e^{-S_0(\varphi)-S_{I}(\varphi)}\, ,
\end{equation}
where
\begin{equation}
S_I(\varphi)\;=\;\frac{1}{2} \, \int d^4x d^4x'\,\varphi(x') V(x,x') \varphi(x)\;.
\end{equation}
In (\ref{eq:pint}) we have neglected a term which, albeit independent of
$\varphi$, may, for a non-rigid motion of the mirror, account for the
excitation of internal degrees of freedom due to the acceleration of the
mirror. These non inertial effects, described in~\cite{inertial}, will not
be considered in the present work since we are here interested in effects
where the vacuum field fluctuations are relevant.

Regarding the form of $V(x,x')$, since both $S^{(0)}_m$ and $S^{(int)}_m$
are localized on the mirror's world-volume,
\begin{equation}\label{eq:ker1}
V(x,x') \;=\;\delta\big(x_3-q_\perp(x_0)\big)\,
\Lambda(x_0,x_\parallel;x_0', x_\parallel') \,
\delta\big(x'_3-q_\perp(x'_0)\big) \;,
\end{equation}
where $\Lambda$ is a -yet unspecified- function, which,  {\em in a system
where the mirror has no parallel motion\/} is assumed to be nonlocal only in
time, namely,
\begin{equation}
\Lambda(x_0,x_\parallel;x_0', x_\parallel') \;=\; \lambda(x_0-x_0') \,
\delta^{(2)}(x_\parallel-x_\parallel') \;.
\label{Lambda}
\end{equation}
The simplifying assumption that $\Lambda(x_0,x_\parallel;x_0', x_\parallel')$ is only
nonlocal in time is tantamount to assuming that its
reflection and transmission coefficients shall be only a function of
frequency, and not on the parallel component of the momentum of the
incident scalar field waves. In other words, the fluctuations of $\psi$ are
only time dependent, they do not propagate in the parallel direction. 

The fact that $\Lambda$ adopts its simplest form in a {\em comoving} system
deserves perhaps a little extra clarification: $\Lambda$ describes the
correlation {\em mediated by the microscopic field\/}, of two
$\varphi$-field fluctuations localized on the world-volume of the mirror.
In a comoving system, that correlation will, since the properties of the
medium are assumed to be independent of time, depend only on the difference
between the times.  

As a concrete example, in the comoving system  one may consider a mirror at
$x_3=0$, composed of decoupled one-dimensional oscillators at each point,
$Q(x_\parallel,x_0)$, each one coupled to $\varphi$ at its respective
position, that is
\begin{eqnarray}
S_m^{(0)}&=&\frac{1}{2}\int dx_0 \int d^2x_\parallel 
\left[\dot Q(x_\parallel,x_0)^2+ \Omega^2Q(x_\parallel,x_0)^2 \right]\nonumber\\
S_{m}^{\rm int}&=& i g\int d^4x \, Q(x_\parallel,x_0) \delta(x_3)
\varphi(x_0,x_\parallel,x_3)\, ,
\end{eqnarray}
where $\Omega$ and $g$ are positive constants.
It is straightforward to check that, in this case, the integration over the
internal degrees of freedom produces an interaction term $S_I$ with a
two-point function $\Lambda(x_0,x_\parallel;x_0', x_\parallel')$  of the
form given in Eq.(\ref{Lambda}), with:
\begin{equation}
\lambda(x_0 - x'_0)  \,=\, \frac{g^2}{ 2 \Omega} e^{- \Omega |x_0-x'_0|} \;.
\end{equation}
In the limit where the oscillators become extremely rigid, $\Omega \to
\infty$, one obtains a time-local interaction: 
\begin{equation}
\lambda(x_0 - x'_0) \,\to\, \big(\frac{g}{\Omega}\big)^2 \, \delta(x_0-x'_0) \;.
\end{equation}

On the other hand, as seen from the Lab system, when the mirror has
parallel motion, the situation does change: $\Lambda$ may depend on
$q_\parallel$, since the microscopic degrees of freedom are not at rest,
and their collective motion may contribute to the correlation of vacuum
field fluctuations between points with a spatial separation.  
The relevant modifications one should implement for parallel motion are discussed
in Section~\ref{sec:friction}. Since the speeds are assumed to be non-relativistic we
shall neglect, for both parallel and normal motions, the time dilation
effects that could also affect the form of $\lambda$.

Coming back to the construction of the model, the final step is to
integrate out the vacuum field itself, 
\begin{equation}\label{eq:geff1}
\Gamma(q_\perp,q_\parallel)\;=\;\frac{1}{2}\,
\log\det(-\partial^2+V)\,=\,\frac{1}{2}{\rm Tr}\log(-\partial^2+V)\;.
\end{equation}
When rotated back to Minkowski spacetime, this effective action
contains information about the reaction force exerted by the vacuum on the
moving mirror, and also on the probabilities of exciting anyone of the two fields
that have been integrated out. 

\subsection{Two mirrors}\label{ssec:twomirrors}

The previous considerations can be generalized to several moving mirrors in
a rather straightforward way. We are here interested in case of just two
mirrors. Besides, since only their relative motion may affect the physical
results, we shall use a reference system, the laboratory frame, where one
of them is at rest. In our conventions, that mirror will be the `left'
($L$) one, located at $x_3=0$, while the other, `right' ($R$) mirror, shall
move rigidly in either the normal or parallel direction. 

The effective action is, in this case, given by an expression that
generalizes (\ref{eq:geff1}):
\begin{equation}
\Gamma(q_L,q_R) =
\frac{1}{2} \log\det(-\partial^2+V_L+V_R) 
=\frac{1}{2}{\rm Tr}\log(-\partial^2+V_L+V_R)\;,
\end{equation}
where $V_L$ ($V_R$) is the kernel for the interaction between the vacuum
field $\varphi$ and the left (right) mirror. We denote by  $q_L (q_R )$ 
the functions that describe either the normal or parallel motion of the left (right) mirror. 

This effective action may be decomposed into three terms which make it 
easier to disentangle the role of each mirror's individual motion from the
contribution of their relative motion. Indeed, defining $\Gamma_I(q_L,q_R)$ by
\begin{equation}
\Gamma(q_L,q_R)\,=\, \Gamma(q_L) \,+\,\Gamma(q_R) \,+\, \Gamma_I(q_L,q_R)\;,
\end{equation}
where the first two terms are just the effective actions corresponding to
each individual mirror (in the absence of the other), while
$\Gamma_I(q_L,q_R)$, by its very definition, captures the part of the effective
action which depends on the influence between the two mirrors. 

In some particular situations it is unnecessary to perform the subtraction
above to find $\Gamma_I$. Indeed, if one knows, on physical grounds, that
each mirror by itself has a trivial effective action, $\Gamma$ equals
(except for irrelevant constants) $\Gamma_I$. This shall be the case for
parallel motion with constant speed.

In the next two sections, we evaluate $\Gamma_I$ for two different
situations, purely normal or purely parallel motion, within the context of
different approximations.
\section{Normal motion}\label{sec:normal}
Under the assumption that the $L$ mirror is static (in the chosen reference
system) and that there is only normal motion for the $R$ mirror:
$q_\perp\neq 0$ and $q_\parallel=0$, the kernels for $V_L$ and $V_R$ become:
\begin{equation}
V_A(x,x') \;=\; \delta\big(x_3-q_A(x_0)\big)\, \lambda(x_0-x_0') \,
\delta\big(x'_3-q_A(x_0')\big)\, \delta^{(2)}(x_\parallel-x'_\parallel), 
\end{equation}
where $A= L, R$, with $q_L= 0$, $q_R = q_\perp\neq 0$.

In this case, an expression for $\Gamma_I$ may be obtained, for example, by a
natural extension of the auxiliary field method used in~\cite{aux} for the
static Casimir effect:
\begin{equation}
\Gamma_I(q_L,q_R) \;=\; \Gamma_I(q_\perp) \;=\; \frac{1}{2} \, \log
\det\big({\mathcal K}\big) \;=\; \frac{1}{2} \, {\rm Tr}\big(\log{\mathcal
K}\big) \;,
\end{equation}
where ${\mathcal K}$ is a $2\times 2$ matrix of kernels,
${\mathcal K}_{AB}(x_0,x_\parallel;x_0',x'_\parallel)$,
\begin{equation}
{\mathcal K}_{AB}(x_0,x_\parallel;x'_0,x'_\parallel) =
\Delta[x_0,x_\parallel, q_A(x_0) ; x_0',x_\parallel', q_B(x_0')]
+ \lambda^{-1}(x_0-x_0') \delta^{(2)}(x_\parallel-x'_\parallel) \,\delta_{AB} \;, 
\end{equation}
where $\Delta$ is the free scalar field propagator in coordinate space 
and $\lambda^{-1}(x_0-x_0')$ is the inverse of $\lambda(x_0-x_0')$, defined
by
\begin{equation}
\int dx_0'' ~ \lambda(x_0-x_0'') ~ \lambda^{-1}(x_0''-x_0')=\delta(x_0-x_0')\, .
\end{equation}
Symmetry under translations in the parallel directions may be exploited
to perform a Fourier transform in those coordinates, so that the problem is
in fact essentially one-dimensional:
\begin{equation}
\Gamma_I(q_L,q_R) \;=\; \frac{\Sigma}{2} \int \frac{d^2k_\parallel}{(2\pi)^2} \, 
{\rm Tr}\big(\log\widetilde{\mathcal K}\big) \;,
\label{onedim}
\end{equation}
where $\Sigma$ is the area of the mirrors and 
\begin{equation}
\widetilde{\mathcal K}_{AB}(x_0,x'_0;k_\parallel) = \int d^2 x_\parallel
e^{- i k_\parallel \cdot (x_\parallel -x'_\parallel)} 
{\mathcal K}_{AB}(x_0,x'_0; x_\parallel-x'_\parallel) ,
\end{equation}
where we have made explicit the fact the kernel can only depend on
differences of parallel coordinates.
Thus,
\begin{equation}
\widetilde{\mathcal K}_{AB}(x_0,x_0';k_\parallel) = 
\widetilde{\Delta}[x_0, q_A(x_0); x_0', q_B(x_0');k_\parallel]
+ {\lambda}^{-1}(x_0-x_0')\delta_{AB}\;.
\label{explicit}
\end{equation}
Note that the trace affects now the time-coordinate continuous indices and
the discrete space of $L$ and $R$ components, but a partial trace over the
parallel space is implemented by the momentum integral. In the rest of this
section we will omit the factor $\Sigma$, i.e. we will compute the
effective actions and forces per unit area of the mirrors.

Following~\cite{Fosco:2007nz}, we compute the effective action assuming
small displacements of the moving mirror around its equilibrium position. 
Expanding the matrix elements $\widetilde{\mathcal K}_{AB}$ in powers of $q_\perp$,
\begin{equation}
\widetilde{\mathcal K}=\widetilde{\mathcal K}_0+
\widetilde{\mathcal K}_1+\widetilde{\mathcal K}_2+....
\end{equation}
where the subindices denote the order of each term (in order to simplify
the notation we omitted the $A,B$ labels of each matrix element).
Therefore,
\begin{equation}
\Gamma_I(q_\perp)\,=\,\frac{1}{2}\int\frac{d^2k_\parallel}{(2\pi)^2}\, {\rm Tr} \log \big[1+{\widetilde{\mathcal
K}}_0^{-1}(\widetilde{\mathcal K}_1+\widetilde{\mathcal K}_2+....)\big]\, , 
\end{equation}
where we omitted a divergent term independent of the position of the
mirrors. Keeping up to quadratic terms in $q_\perp(\tau)$ we have
\begin{equation}
\Gamma_I(q_\perp)\,=\, \frac{1}{2}\int\frac{d^2k_\parallel}{(2\pi)^2}\, {\rm Tr} \big[\widetilde{\mathcal K}_0^{-1}(\widetilde{\mathcal K}_1 +\widetilde{\mathcal K}_2)
-\frac{1}{2}(\widetilde{\mathcal K}_0^{-1}\widetilde{\mathcal K}_1)^2
\big]\, .
\label{Gammapert}
\end{equation}

We now present some details of the evaluation of  $\Gamma_I$, and of some
physical observables derived from it. We do that mostly in $3+1$
dimensions; the special case of $1+1$ dimensions is considered at the end.

First we consider the form of the kernels $\widetilde{\mathcal K}_{iAB}$,
$i=0,1,2$.  
From (\ref{explicit}), we see that the matrix elements of the operator
$\widetilde{\mathcal K}$ read, to the lowest order,
\begin{equation}
\widetilde{\mathcal K}_{0AB}(x_0,x_0';k_\parallel)=\int\frac{d\omega}{2\pi}e^{i\omega(x_0-x_0')}
\widetilde{\widetilde{\mathcal K}}_{0AB}(\omega, k_\parallel)\, ,
\end{equation}
where 
\begin{equation}
\widetilde{\widetilde{\mathcal K}}_{0AB}(\omega,k_\parallel)=\frac{1}{\widetilde\lambda(\omega)}\delta_{AB}
+\tilde\Delta_{0AB}(\omega, k_\parallel)\,
\end{equation}
and
\begin{equation}
\tilde\Delta_{0}(\omega,k_\parallel)=\frac{1}{2\sqrt{\omega^2+k_\parallel^2 +m^2}}
\left( \begin{array}{ccc}
1 & e^{-a\sqrt{\omega^2+ k_\parallel^2 + m^2}}  \\
 e^{-a\sqrt{\omega^2+ k_\parallel^2 + m^2}}  & 1 \end{array} \right)\, .
\end{equation}
In the above equations we have introduced the Fourier transform
\begin{equation}
\widetilde{\lambda}(\omega) = \int d\tau e^{-i \omega \tau} 
\lambda(\tau)\, .
\end{equation}

To first order in $q_\perp(x_0)$ we have
\begin{equation}
\widetilde{\mathcal K}_{1LR}(x_0,x_0';k_\parallel) =
\widetilde{\mathcal K}_{1RL}(x_0',x_0;k_\parallel)
=-\frac{q_\perp(x_0)}{2}\int\frac{d\omega}{2\pi}e^{i\omega(x_0-x_0')}e^{-a\sqrt{\omega^2+
k_\parallel^2 + m^2}}\, , \label{K1}
\end{equation}
while $\widetilde{\mathcal K}_{1LL}=\widetilde{\mathcal K}_{1RR}=0$.

The second order non-vanishing matrix elements are
\begin{eqnarray}
\widetilde{\mathcal
K}_{2RR}(x_0,x_0',k_\parallel)&=&\frac{1}{4}[q_\perp(x_0)-q_\perp(x_0')]^2
\int
\frac{d\omega}{2\pi}e^{i\omega(x_0-x_0')}\sqrt{\omega^2+k_\parallel^2 +m^2}
\nonumber \\
\mathcal K_{2LR}(x_0,x_0',k_\parallel)
&=&\frac{1}{4}q_\perp(x_0)^2\int
\frac{d\omega}{2\pi}e^{i\omega(x_0-x_0')}\label{K2} e^{-a\sqrt{\omega^2+k_\parallel^2 +m^2}}\sqrt{\omega^2+k_\parallel^2 +m^2} 
\end{eqnarray}
and $\widetilde{\mathcal K}_{2RL}(x_0',x_0;k_\parallel)=\widetilde{\mathcal K}_{2LR}(x_0,x_0';k_\parallel)$.
\subsection{Linear term in $q$: the static Casimir force}
As a first result, we obtain the form of the linear term in the effective
action. It is given by 
\begin{eqnarray}
\Gamma_1(q_\perp)&=&\frac{1}{2}\int\frac{d^2k_\parallel}{(2\pi)^2}\, {\rm
Tr} [\widetilde{\mathcal K}_0^{-1}\widetilde{\mathcal K}_1]
=\frac{1}{2}\int \frac{d^2k_\parallel}{(2\pi)^2}\int dx_0
dx_0'\left[\widetilde{\mathcal
K}_{0LR}^{-1}(x_0,x_0',k_\parallel)\widetilde{\mathcal
K}_{1RL}(x_0,x_0';k_\parallel) 
\right. \nonumber \\
&+& \left. \widetilde{\mathcal K}_{0RL}^{-1}(x_0,x_0';k_\parallel)
\widetilde{\mathcal K}_{1LR}(x_0,x_0';k_\parallel)\right]
\, .
\end{eqnarray}
The inverse of the free kernel, $\widetilde{\mathcal K}_0^{-1}$, can be
exactly evaluated because $\widetilde{\mathcal K}_0$  is diagonal in
frequency space. Its Fourier transform is given by 
\begin{equation}
\widetilde{\widetilde  {\mathcal K}}_{0}^{-1}(\omega,k_\parallel)=
\frac{1}{c^2(\omega,k_\parallel)-b^2(\omega, k_\parallel)} 
\left( \begin{array}{ccc}
c(\omega,k_\parallel)& -b(\omega,k_\parallel)  \\
 -b(\omega,k_\parallel) &c(\omega,k_\parallel) \end{array} \right)\, ,
 \label{invK0}
\end{equation} 
where we have introduced: 
 \begin{eqnarray}
 b(\omega,k_\parallel)&=& \frac{e^{-a\sqrt{\omega^2+k_\parallel^2 +m^2}}}
{2\sqrt{\omega^2+k_\parallel^2 +m^2}} \nonumber\\  
c(\omega,k_\parallel)&=&
\frac{1}{\widetilde\lambda(\omega)} +
\frac{1}{2\sqrt{\omega^2+k_\parallel^2 +m^2}} \;.
\label{invK0bis}
\end{eqnarray}
Therefore, we obtain for the linear part of the effective action
\begin{equation}
\Gamma_1(q_\perp)=-\frac{1}{2}\int dx_0\, q_\perp(x_0)\int
\frac{d\omega}{2\pi}\int\frac{d^2k_\parallel}{(2\pi)^2}\frac{1}{\sqrt{\omega^2+k_\parallel^2 +m^2}} 
\frac{e^{-2a\sqrt{\omega^2+k_\parallel^2
+m^2}}}{\left(\frac{1}{\widetilde\lambda(\omega)}+\frac{1}{2\sqrt{\omega^2+k_\parallel^2
+m^2}}\right)^2
-\frac{e^{-2a\sqrt{\omega^2+k_\parallel^2 +m^2}}}{4(\omega^2+k_\parallel^2 +m^2)}   } .
\end{equation}
This result has a clear interpretation. The functional variation of the
effective action with respect to  $q_\perp(x_0)$ gives the force on the
left mirror, which in this  approximation is time independent and given by
the usual static Casimir force between thin mirrors characterized by a
function $\widetilde\lambda(\omega)$.  The usual Casimir force
for Dirichlet mirrors is recovered in the $m=0$ and
$\widetilde\lambda=\infty$ limit.

\subsection{Quadratic term: decay of the vacuum and dissipative force on
the moving mirror}\label{sec:quad}
The linear term in the effective action does not contain information about
the back-reaction of the created particles on the motion of the mirror.
Indeed, we have seen in the previous subsection that it has the form
\begin{equation}
\Gamma_1(q_\perp)=\int dx_0 \,q_\perp(x_0) F_C
\end{equation} 
where $F_C$ is the (time-independent) static Casimir force between
imperfect mirrors separated by a distance $a$.  On general grounds, we
expect that, to second order,
\begin{equation}
\Gamma_2(q_\perp)= \frac{1}{2} \, \int dx_0\int dx_0' \,q_\perp(x_0)
F(x_0-x_0') q_\perp(x'_0) \,.
\label{Gamma2gen}
\end{equation}
This nonlocal effective action contains (in its kernel) information on both 
the dissipative force on the mirror and on the probability of creating $\varphi$
excitations. 

To calculate $F$, we note that, from (\ref{onedim}) and (\ref{Gammapert}):
\begin{equation}
\Gamma_2(q_\perp)=\frac{1}{2}\int\frac{d^2k_\parallel}{(2\pi)^2}\, {\rm Tr} \big[
\widetilde{\mathcal K}_0^{-1}\widetilde{\mathcal K}_2
-\frac{1}{2}(\widetilde{\mathcal K}_0^{-1}\widetilde{\mathcal K}_1)^2
\big]\equiv \Gamma_2^{(1)}+\Gamma_2^{(2)}\, .
\label{Gamma2pert}
\end{equation} 
The first contribution can be written more explicitly as
\begin{equation}
\Gamma_2^{(1)}=\frac{1}{2}\int dx_0 \int dx_0'
\int\frac{d^2k_\parallel}{(2\pi)^2}\widetilde{\mathcal
K}_{0AB}^{-1}(x_0,x_0';k_\parallel)\widetilde{\mathcal
K}_{2BA}(x_0',x_0;k_\parallel) \end{equation}
The term proportional to $\widetilde{\mathcal
K}_{0RR}^{-1}\widetilde{\mathcal K}_{2RR}$ carries a dependence in $a$, and
it describes, when $a \to \infty$, the effective action for the moving $R$
mirror in the absence of the $L$ mirror: we already computed this
single-mirror term in~\cite{Fosco:2007nz}, for the particular case of a constant 
$\widetilde{\lambda}$. 

For a finite $a$, and after discarding terms which renormalize the mass of
the mirror, we see that:
\begin{equation}\label{eq:g21}
\Gamma_2^{(1)}=\frac{1}{2} \,
\int dx_0 \int dx_0' \, q_\perp(x_0) F^{(1)}(x_0-x_0') q_\perp(x'_0) 
\end{equation}
where the Fourier transform of $F^{(1)}$ is
\begin{equation}
\widetilde{F^{(1)}}(\omega) \,=\, -\frac{1}{4 \pi} \int d\nu \,
\int\frac{d^2k_\parallel}{(2\pi)^2}
\frac{ c(\nu + \omega) \sqrt{\nu^2 + k_\parallel^2 +
m^2}}{c^2(\nu+\omega,k_\parallel) - b^2(\nu+\omega,k_\parallel)} \;. 
\end{equation}

On the other hand, the terms  proportional to $\widetilde{\mathcal
K}_{0LR}^{-1} \widetilde{\mathcal K}_{2RL}$ and $\widetilde{\mathcal
K}_{0RL}^{-1}\widetilde{\mathcal K}_{2LR}$ are of the form $\Delta M^2\int
dx_0 q^2(x_0)$ and produce a (finite) renormalization of the mass of the
mirror, which we again discard from our expression.

We now evaluate the second contribution to (\ref{Gamma2pert}), which reads
\begin{equation}
\Gamma_2^{(2)}(q_\perp)=-\frac{1}{4}\int d\tau_1d\tau_2d\tau_3d\tau_4
\,\widetilde{\mathcal
K}_{0AB}^{-1}(\tau_1,\tau_2;k_\parallel)\,\widetilde{\mathcal
K}_{1BC}(\tau_2,\tau_3;k_\parallel) \,\widetilde{\mathcal
K}_{0CD}^{-1}(\tau_3,\tau_4;k_\parallel)\,\widetilde{\mathcal K}_{1DA}
(\tau_4,\tau_1;k_\parallel)\, ,
\end{equation}
and again may be put under a form similar to (\ref{eq:g21}):
\begin{equation}\label{eq:g22}
\Gamma_2^{(2)}=\frac{1}{2} \,
\int dx_0 \int dx_0' \; q_\perp(x_0) F^{(2)}(x_0-x_0') q_\perp(x'_0) 
\end{equation}
where now
\begin{eqnarray}
\widetilde{F^{(2)}}(\omega)&=& -\frac{1}{8\pi} \int
\frac{d^2k_\parallel}{(2\pi)^2} \int d\nu  \,
\frac{1}{c^2(\omega+\nu,k_\parallel)-
b^2(\omega+\nu,k_\parallel)}  \frac{1}{c^2(\nu,k_\parallel)-
b^2(\nu,k_\parallel)}\nonumber\\ &\times &
\left[c(\omega+\nu,k_\parallel)c(\nu,k_\parallel) e^{-2 a\sqrt{(\omega +
\nu)^2+ k_\parallel^2 + m^2}}+
b(\omega+\nu,k_\parallel)b(\nu,k_\parallel)e^{-a\sqrt{(\omega +\nu)^2+ k_\parallel^2 +
m^2}}e^{-a\sqrt{\nu^2+ k_\parallel^2 + m^2}} \right]\,.
\end{eqnarray}
Thus, the form of $\Gamma_2$ shall be:
\begin{equation}\label{eq:g2}
\Gamma_2=\frac{1}{2} \,
\int dx_0 \int dx_0' \; q_\perp(x_0) F(x_0-x_0') q_\perp(x'_0) 
\end{equation}
where $F = F^{(1)}+F^{(2)}$. 

This is the main result of this section. It gives, to second order in the
departure from the normal moving mirror's average position, the Euclidean effective
action. 
At the same order, the {\it in-out} effective action becomes
\begin{equation}
\Gamma_{2, {\rm in-out}}(q_\perp)=\frac{1}{2}\int dx_0\int dx_0'
\,q_\perp(x_0) F_{{\rm in-out}}(x_0-x_0') q_\perp(x'_0)\, , 
\label{Gamma2geninout}
\end{equation}
where $F_{\rm{in-out}}$ is the continuation of the kernel to Minkowski
spacetime, using Feynman's prescription to avoid the poles.  Moreover, the
dissipative force on the moving mirror is given by 
\begin{equation}
 F(x_0)=\int dx'_0 F_{\rm ret}(x_0-x'_0) q(x'_0)\, ,
\end{equation}
where $F_{\rm ret}$ is the (retarded) continuation to Minkowski spacetime. 
\subsection{$1+1$ dimensions}
We will now present some explicit evaluations of the effective action for
the $1+1 $ dimensional case. Its formal expression can be obtained in a
simple fashion from the previously derived  ones.  
Indeed, beginning from the computation of the kernel $F(x_0-x_0')$ in $3+1$ dimensions
one should omit the parallel coordinates $x_\parallel$ and the corresponding 
integration in the $k_\parallel$ momenta. 

The result for $F^{(1)}$ is 
\begin{equation}
\widetilde{F^{(1)}}(\omega) \,=\, -\frac{1}{4 \pi} \int d\nu \,
\frac{ c(\nu + \omega) \sqrt{\nu^2 + m^2}}{c^2(\nu+\omega) - b^2(\nu+\omega)} 
\end{equation}
while for $F^{(2)}$: 
\begin{equation}
\widetilde{F^{(2)}}(\omega)= -\frac{1}{8\pi}\int d\nu \,
\frac{c(\omega+\nu)c(\nu) e^{-2a\sqrt{(\omega+\nu)^2+m^2}}+
b(\omega+\nu)b(\nu)e^{-a\sqrt{(\omega +\nu)^2+m^2}}
e^{-a\sqrt{\nu^2+m^2}}}{(c^2(\omega+\nu)-b^2(\omega+\nu))(c^2(\nu)-b^2(\nu))}
\;.
\label{F1+1}
\end{equation}

In the particular case of imperfect mirrors such that
$\tilde{\lambda}(\omega) = \zeta |\omega|$ (see next section and Appendix
A) and $m=0$, the expressions above simplify to: 
\begin{equation}
\widetilde{F^{(1)}}(\omega) \,=\, -\frac{1}{2\pi} \int d\nu \,
|\nu + \omega| \, |\nu| \; \frac{ \chi }{\chi^2 - e^{- 2 a |\nu+\omega|}}
\label{F1chi}
\end{equation}
and
\begin{equation}
\widetilde{F^{(2)}}(\omega) = -\frac{1}{2\pi}
\int d\nu \, e^{-2 a |\omega + \nu |}\frac{|\omega + \nu | |\nu|}{\left(\chi^2 - e^{-2 a |\omega + \nu |}\right)} \frac{\left(\chi^2 + e^{-2 a |\nu |}\right)}{\left( \chi^2 - e^{-2 a | \nu |}\right)},
\label{F2chi}
\end{equation} 
where we have set $\chi = (2+\zeta)/\zeta$. The perfect mirror case is
obtained by taking the $\zeta \to \infty$ limit: 
\begin{equation}
\widetilde{F^{(1)}}(\omega) \to \widetilde{F_\infty^{(1)}}(\omega)  
\,=\, -\frac{1}{2\pi} \int d\nu \,
|\nu + \omega| \, |\nu| \; \frac{1}{e^{2 a |\nu+\omega|} - 1}
\end{equation}
\begin{equation}
\widetilde{F^{(2)}}(\omega) \to \widetilde{F_\infty^{(2)}}(\omega)  
= -\frac{1}{2\pi} \int d\nu \, 
e^{-2 a |\omega + \nu |}
\frac{|\omega + \nu | |\nu|}{\left(1 - e^{-2 a |\omega + \nu |}\right)} 
\frac{\left(1 + e^{-2 a |\nu |}\right)}{\left( 1 - e^{-2 a | \nu |}\right)},
\end{equation}
where we have subtracted the (already known)
contribution corresponding to a single mirror from ${\widetilde
F^{(1)}}(\omega)$.

Thus, for perfect mirrors, we have:
\begin{eqnarray}
\widetilde{F_\infty}(\omega) &=& \widetilde{F_\infty^{(1)}}(\omega) +
\widetilde{F_\infty^{(2)}}(\omega) \nonumber\\
&=&
\frac{\vert\omega\vert^3}{12\pi}-\frac{\omega^2\pi}{6a^3}(1+\frac{\omega^2
a^2}{\pi^2})\sum_{n\geq 1}\frac{1}{\omega^2+\frac{n^2\pi^2}{a^2}}\;.
\end{eqnarray}

This form of the kernel is useful to understand its general structure and to  perform the analytic continuations. 
Indeed, following the
procedure outlined in Ref.\cite{Fosco:2007nz}, we can use the integral representation
\begin{equation}\vert\omega\vert ^3 =
\frac{2\omega^4}{\pi}\int_0^{+\infty}dz \frac{1}{\omega^2 +
z^2},\end{equation}
so the full kernel is written in terms of the massive $0+1$ dimensional Euclidean propagator $(\omega^2+M^2)^{-1}$. 
Therefore the $in-out$ kernel for the effective action
can be obtained  through the substitution $\omega^2\rightarrow
-\omega^2+i\epsilon$. Moreover, the force on the mirror can be obtained by replacing the Euclidean propagator by the retarded one, i.e. $\omega^2\rightarrow -(\omega+i\epsilon)^2$.  We see that the real time kernels have poles for the particular frequencies $\omega_n=n\pi/a$. These poles
indicate the well known breakdown of the perturbative calculation when the mirror oscillates at such resonant frequencies. One can check that the analytic continuations
coincide with the results previously found by other authors \cite{perf1+1}. In particular, the retarded kernel reads
\begin{equation}
F_{\rm ret}(x_0)= \frac{\delta'''(x_0)}{12\pi}-\frac{\pi}{6a^2}\theta(t)\sum_{n\geq 0}\delta'(x_0-2 na)-\frac{1}{6\pi}\theta(t)\sum_{n\geq 0}\delta'''(x_0-2 na)\, .
\end{equation}

This analysis can be generalized to arbitrary (bigger than $1$) values of
$\chi$. The kernels in Eqs. \ref{F1chi} and \ref{F2chi} can be computed
analytically. However, the resulting expressions are rather complicated and
not very illuminating. In that respect, it is more useful to consider the
case of quasi transparent mirrors: $\chi >> 1$ ($\zeta << 1$). The
leading terms in an expansion in powers of $\chi^{-1}$ are:
\begin{eqnarray}
\widetilde{F^{(1)}}(\omega) &=& -\frac{1}{2\pi\chi} \int d\nu \,
|\nu + \omega| \, |\nu| \; (1 +  \frac{e^{- 2 a |\nu+\omega|}}{\chi^2})+\dots \nonumber\\
\widetilde{F^{(2)}}(\omega) &=& -\frac{1}{2\pi\chi^2}\int d\nu \,|\omega + \nu | |\nu| e^{-2 a |\omega + \nu |}+\dots
\;.
\end{eqnarray} 
Therefore:
\begin{equation}
\widetilde{F}(\omega)\,=\, -\frac{1}{6\pi\chi}\vert\omega\vert^3
-\frac{1}{4\pi a^3\chi^2}[e^{-2a\vert\omega\vert}(1+a\vert\omega\vert)+a\vert\omega\vert]+\dots
\end{equation}
The leading term in the expansion does not depend on $a$, and equals the
kernel of a single imperfect moving mirror.  Being proportional  to $\vert
w\vert^3$ (or $\delta'''(x_0)$ in its retarded Fourier transform), it is
similar to that of the single perfect moving mirror case. Its form
could have been anticipated by using dimensional analysis; indeed, for this
particular form of $\tilde\lambda(\omega)$, $\zeta$ is dimensionless, and
therefore  one does not have any additional dimensionful constant in the theory. 

Using the integral representation
\begin{equation}  \vert\omega\vert e^{-2\vert\omega\vert  a}=\frac{2\omega^2}{\pi}\int_z^\infty dz \frac{e^{2 i  z a}}{\omega^2+z^2}
\, ,\end{equation}
one can show that the $1/\chi^2$-retarded contributions are proportional to
$\delta'(x_0), \delta(x_0-2a), $ and
$\delta'(x_0-2a)$. Higher order corrections always involve the $\delta$-function
and its derivatives evaluated at $x_0-2na$. Once more, this is a characteristic of 
the specific form used for $\tilde\lambda(\omega)$: as the corresponding
reflection coefficient has a constant phase,  the mirrors do not introduce
any delay in a reflecting wave packet. Thus the time of flight between the two mirrors 
does not depend on $\chi$, and equals the one for perfect mirrors
($\chi=1$).

\section{Sidewise motion}\label{sec:friction}
Here, the $L$ mirror is again at rest and lying on the $x_3=0$ plane, while the 
$R$ mirror, at a constant distance of the first, is at $x_3=a$, moves along
$x_1$: $x_1= q_\parallel(x_0)$.

It is worth to mention that, in this case, one should expect forces
only on an {\it imperfect} mirror. Indeed, there can be no force for a {\it
perfect} mirror, since in such a case the vacuum field will satisfy
Dirichlet boundary conditions, irrespective of its state of motion, and the
effective action $\Gamma_I$ becomes proportional to the Casimir energy for
two perfect mirrors. 
However, on an imperfect mirror the boundary conditions will, as we shall
see,  depend on the state of motion. Therefore, there will be excitations
both of the vacuum field and of the microscopic degrees of freedom of the
mirror.  

Since the $L$ mirror is at rest, the form of its interaction kernel is,
simply 
\begin{equation}
V_L(x,x')=\delta(x_3)\,\lambda(x_0-x'_0)\delta^{(2)}(x_\parallel-x'_\parallel) \,\delta(x'_3)\,,
\end{equation}
while the $R$ has a similar form only in a comoving system: 
\begin{equation}
V_R^{(0)}(x,x')=\delta(x_3-a)\,\lambda(x_0-x'_0)\delta^{(2)}(x_\parallel-x'_\parallel)
\,\delta(x'_3-a)\,,
\end{equation}
where the $(0)$ reminds us of the fact that it corresponds to a system
where $R$ is at rest.
The key point is that, if there is relative motion there is no system where
both mirrors are at rest.  The  interaction is spatially local in the rest
frame, when the mirror is moving becomes spatially nonlocal, since it shall
link points that are now connected by the evolution of the mirror:
\begin{equation}
\delta(x_1-x'_1)\rightarrow \delta[x_1-x'_1-q_\parallel (x_0)+q_\parallel (x'_0)]\,.
\end{equation}
Thus, the form of the $V_R$ kernel is:
\begin{equation}
V_R(x,x')=\delta(x_3-a)\,\lambda(x_0-x'_0)\, \delta[x_1-x'_1-q_\parallel (x_0)+q_\parallel (x'_0)]\,\delta(x_2-x'_2) 
\,\delta(x'_3-a)\,.
\end{equation}

In the equations above  $q_\parallel (x_0)$ describes the rigid translation of the
mirror; $v(x_0)=\dot q_\parallel (x_0)$ will be used for its instantaneous velocity.
It is clear that $V_R$ will, in general, acquire a  non trivial time
dependence, which shall be the origin of the dissipative effects described
below. 

In what follows, we compute the effective action for different kinds of
motions (of the R mirror).  

\subsection{Constant velocity (Casimir friction)} 
As shown several years ago by Pendry~\cite{Pendry97}, two flat surfaces
characterized by different dielectric functions experience a friction force
when sheared parallel to their interface, even at a constant velocity. This
effect is produced by the exchange of virtual photons, that "see" a
different reflection coefficient on each surface, due to the non vanishing
relative velocity.  Recently, there has been some debate about the reality
of this effect, see for instance~\cite{debate}. This in turn has triggered
additional works \cite{others} that confirmed the frictional Wan der Walls forces between
atoms, and atom near a surface, and also between surfaces.  We will now
show that dissipative effects are also present for the thin mirrors
considered in this paper, as  long as the effective action for the vacuum
field is nonlocal in time.

Note that, in this constant velocity case, it is entirely equivalent to
consider $\Gamma_I$ or $\Gamma$,  since it is only the relative motion what may
produce friction. The effective action corresponding to each isolated
mirrors is trivial, as one can guess intuitively, and also by an explicit
calculation. 

To proceed, we expand the effective action perturbatively in $\lambda$. To
lowest nontrivial order,
\begin{equation}
\Gamma_I \approx -\frac{1}{2} {\rm Tr}\left(\frac{1}{-\partial^2}V_L\frac{1}{-\partial^2}V_R\right)   \, ,
\end{equation}
or, in Fourier space,
\begin{equation}
\Gamma_I \approx-\frac{1}{2}\int \frac{d^4p}{(2\pi)^4}
\frac{d^4q}{(2\pi)^4}\widetilde G(p) \widetilde V_L(p,q)\widetilde
G(q)\widetilde V_R(q,p)
\end{equation}
where:
\begin{eqnarray}
\widetilde G(p)&=&\frac{1}{p^2+m^2}\nonumber\\
\widetilde V_{A}(p,q)&=&\int d^4x\, d^4y\, e^{-ipx +iqy} \, V_A(x,y)\, .
\end{eqnarray}
For the particular case of constant velocity $\dot q_\parallel=v$, assuming that the vacuum
field is massless, and, as always, that both mirrors are identical  at rest,
\begin{equation}
\Gamma\approx\frac{T\Sigma}{64\pi^3}\int d^3p\frac{e^{-2a\sqrt{p_0^2+p_1^2+p_2^2}}}{p_0^2+
p_1^2+p_2^2}\widetilde\lambda(p_0)\widetilde\lambda(p_0+p_1 v)\,
\label{Gammapertv}
\end{equation}
which is the main result of this section. From it we can derive several
interesting conclusions:  on the one hand, the derivative of the effective
action with respect to $a$, yields the usual attractive Casimir force between
the mirrors, if evaluated perturbatively for `highly transparent' mirrors.

Equation~(\ref{Gammapertv}) shows, explicitly, that this force depends on the
velocity of the right mirror. This is to be expected, since  the motion of
the mirror changes its reflection coefficient, and therefore affects its
interaction with the vacuum field. On the other hand, in general $\Gamma$
will have an imaginary part when rotated back from Euclidean to Minkowski
spacetime. This is the signal of frictional forces between mirrors.
Moreover, one can prove that in order to have friction between the mirrors
it is necessary that $\widetilde\lambda(p_0)$  have a  non-analyticity.
In other words, the microscopic degrees of freedom in the mirror should
induce a non local interaction for the vacuum field. Indeed, when
$\widetilde\lambda(p_0)$ is analytic, then $\lambda(x_0)$ can be written as
a linear combination of the delta function and its derivatives, so the
interaction $S_I$ can be expanded in local terms involving time-derivatives
of the vacuum field. For this particular interaction, the effective action
(\ref{Gammapertv}) will be analytic in $v$, and only even powers of $v$
will contribute to the final results. These terms will not produce an
imaginary part when the Euclidean velocity is rotated to Minkowski
spacetime $v\rightarrow iv$.
 
As an example to illustrate this point, for a constant velocity, we will
assume that $\widetilde \lambda(p_0)=\zeta \vert p_0\vert$.  
This is a relevant non-analytic interaction since, as shown in
Appendix A, it describes a zero-width mirror where the vacuum field
propagates with velocity $1/\sqrt{1+\zeta}$; in other words,  $1+\zeta$ plays the role
of a  dielectric constant.

The $v$-dependent part of the effective action becomes, in this case,
\begin{equation}
\Gamma\approx\frac{T\Sigma\zeta^2}{576\pi^3}\frac{\vert v\vert^3}{a^3} + O(v^4)\, ,
\end{equation}
which has a non vanishing imaginary part when continued to Minkowski spacetime.
\begin{equation}
{\rm Im}\Gamma_{{\rm in-out}}\approx\frac{T\Sigma\zeta^2}{576\pi^3}\frac{\vert
v\vert^3}{a^3} + O(v^4)\, .
\end{equation}

\subsection{Non-constant velocity}
In this section we compute the form of the effective action for a more general
motion of the right mirror.
We will follow a similar approach to the one 
in Section~\ref{sec:normal}, and compute the effective action in an expansion in
powers of $q_\parallel (x_0)$. The difference is, in this case, that $q_\parallel$ describes
the motion of the $R$ mirror along the $x_1$-direction.

We shall assume that the mirror is always close to its equilibrium
position, and therefore expand the  expression for the $V_R$ in powers of
$q(x_0)$; namely, $V_R=V_R^{(0)}+V_R^{(1)}+V_R^{(2)}+...$, where:
 \begin{eqnarray}
 V_R^{(0)}(x,x')&=&\delta(x_3-a)
\lambda(x_0-x'_0)
\delta^{(2)}(x_\parallel-x'_\parallel) \delta(x'_3-a)\nonumber\\
V_R^{(1)}(x,x')&=&-(q_\parallel(x_0)-q_\parallel(x'_0))\lambda(x_0-x'_0)\delta'(x_1-x'_1)\delta(x_2-x'_2)
  \delta(x_3-a)\delta(x'_3-a)\nonumber\\
   V_R^{(2)}(x,x')&=&\frac{1}{2}(q_\parallel(x_0)-q_\parallel(x'_0))^2
\lambda(x_0-x'_0)\delta''(x_1-x'_1)\delta(x_2-x'_2)  \delta(x_3-a)\delta(x'_3-a)\, .
  \end{eqnarray}

Therefore,  there is then an expansion for $\Gamma$,  $\Gamma_I=\Gamma^{(0)}+\Gamma^{(1)}
+\Gamma^{(2)}+...$, with
\begin{equation}
\Gamma^{(1)}=\frac{1}{2}{\rm Tr}[(-\partial^2+V_L+V_R^{(0)})^{-1}V_R^{(1)}]
\end{equation}
and
\begin{eqnarray} 
 \Gamma^{(2)}&=&\frac{1}{2}{\rm Tr}[(-\partial^2+V_L+V_R^{(0)})^{-1}V_R^{(2)}]
 -\frac{1}{4}{\rm Tr}\left[\left((-\partial^2+V_L+V_R^{(0)})^{-1}V_R^{(1)}\right)^2\right]  \nonumber\\
 &\equiv&\Gamma^{(2,1)}+\Gamma^{(2,2)}\, .
 \end{eqnarray}
 
 In is straightforward to see that $\Gamma^{(1)}=0$; that is, that there is no
linear term in the velocity. Introducing the notation 
 \begin{equation}
 {\cal G}=(-\partial^2+V_L+V_R^{(0)})^{-1}
 \label{G0}
 \end{equation}
 and denoting by $\widetilde {\cal G}$ its Fourier transform, one sees that
 \begin{equation}
 \Gamma^{(2,1)}\;=\; \Sigma \, \int dk_0 \widetilde q(-k_0) \widetilde \Pi^{(2,1)}(k_0)\widetilde
 q(k_0)\end{equation}
 where
 \begin{equation}
 \widetilde\Pi^{(2,1)}(k_0)\;=\;\frac{1}{32\pi^4} \, \int d\omega dk_1 dk_2 \, k_1^2\, 
\widetilde\lambda(\omega-k_0) \widetilde {\cal G}(\omega,k_1,k_2,a,a)\, .
 \end{equation}
 
 The evaluation of $\Gamma^{(2,2)}$ is straightforward, although
algebraically more involved. The final result can be written as
 \begin{equation}
 \Gamma^{(2,2)}=\Sigma\int dk_0\widetilde q(k_0)\widetilde q(-k_0)\widetilde
 \Pi^{(2,2)}(k_0)\, ,
 \end{equation}
 where
 \begin{eqnarray}
 \widetilde \Pi^{(2,2)}(k_0)&=&\frac{1}{64\pi^4}\int dl_0
\Big[2\widetilde\lambda(l_0)\widetilde\lambda(k_0+l_0)
-\widetilde\lambda(k_0+l_0) \widetilde\lambda(k_0+l_0)
-\widetilde\lambda(l_0) \widetilde\lambda(l_0) \Big]
\nonumber\\
 &\times& \int dk_1dk_2 k_1^2
  \widetilde {\cal G}(k_0+l_0,k_1,k_2,a,a)\widetilde {\cal G}(l_0,k_1,k_2,a,a)\, .
  \end{eqnarray}
 
Finally, we note that,  as at the end of Section \ref{sec:quad}, the force on the mirror and 
the imaginary part of the {\it in-out} effective action can be obtained by appropriate continuations
to Minkowski spacetime of the Euclidean form factor $\Pi^{(2,1)}+\Pi^{(2,2)}$.
Moreover, there are dissipative effects even for a single mirror sliding with non-vanishing acceleration
\cite{barton sideways}.

\section{Conclusions}\label{sec:concl}
In this paper we have presented a functional approach to compute the
dissipative effects on imperfect moving mirrors. A crucial  point in our
approach is that the interaction of the vacuum field with the mirror is
modeled by a {\it nonlocal} term in the action of the quantum field. This
kind of term is generated by the interaction of the field with the
internal degrees of freedom of the mirror, in the lowest non trivial order
in an expansion in powers of the field.

The consideration of the quantum theory associated to the degrees of
freedom of the {\it whole} system (vacuum field and mirrors) allowed us to
compute both the dissipative effects coming from the excitation of the
vacuum field, and those coming from the excitation of the internal degrees
of freedom of the mirror. Moreover, we have seen that it is particularly
convenient to work in imaginary time (Euclidean spacetime), in order to
obtain the probability of excitation of the system and the dissipative
force on the moving mirror from adequate analytic continuations of the
effective action to  Minkowski spacetime. 
 
For the sake of simplicity, we have considered a simplified model with a quantum scalar
field, coupled to  flat thin mirrors which undergo  rigid motion. We have also
assumed that the interaction term is nonlocal in time, and local in the
spatial coordinates of the mirror, as it happens when the internal degrees of
freedom can be described by a set of independent harmonic oscillators.
Under these assumptions, we have been able to analyze several interesting
situations that involve two mirrors in normal or parallel relative motion.
In the first case, we computed the effective action of the system for a
moving mirror near a static mirror, perturbatively  in the departure from
the equilibrium position. In the second case, we have shown that there is a
dissipative force between the mirrors  when the sliding motion has constant
velocity. We have seen that the temporal nonlocality is a  necessary
condition for non-vanishing friction. Moreover, using again a perturbative
approach, we have obtained a general expression for the effective action as
a functional of the position of the mirror. As a particular case, we
pointed out that there is dissipation even for a single mirror with sliding
accelerated motion.
 
 The approach described here can be generalized in several directions. On
the one hand, one should consider the more realistic case of the
electromagnetic field. On the other, one could consider more general
nonlocal interactions,  induced by internal degrees of freedom that
propagate inside the mirror. For instance, it would be interesting to
analyze the dynamical Casimir effect for graphene sheets. In this case, the
internal degrees of freedom can be described with massless Dirac fields
\cite{graph}, propagating at a velocity $v_F<<c$ inside the mirror,  and
will produce an effective interaction that is both temporally and spatially
nonlocal. 

\section*{Acknowledgements}
C.D.F. thanks CONICET, ANPCyT and UNCuyo for financial support. The work of
F.D.M. and F.C.L was supported by UBA, CONICET and ANPCyT.

\section*{Appendix A} 
In this Appendix we show that, when the interaction between the internal
degrees of freedom of a zero-width mirror and the vacuum field is described
by the non-analytic function $\tilde\lambda(p_0)=\zeta\vert p_0\vert$, the
the vacuum field propagates with velocity $1/\sqrt{1+\zeta}$ on the mirror.

We consider then a mirror static at $x_3=0$, and to study the  propagation
of $\varphi$ modes on that plane, we add a coupling to an external source
$J$, which is localized on the mirror: $J(x) = j(x_0,x_1,x_2) \delta(x_3)$.  
 
Then, by integrating out the vacuum field in the resulting theory, 
one finds the generating functional:
\begin{equation}
{\mathcal Z}(j) \,=\, \int {\mathcal D} \varphi \; 
e^{-S(\varphi) + i \int d^4x J(x) \varphi(x)}
\end{equation}
whence we can obtain the propagator for $\varphi$ exactly on the plate, by
taking derivatives of ${\mathcal Z}(j)$ with respecto to $j$. 

The integral over the vacuum field is of course Gaussian, and its result is
\begin{equation}
\langle \varphi(x_0,x_1,x_2)  \varphi(x'_0,x'_1,x'_2) 
\rangle \,=\, 
\int \frac{d\omega}{2\pi}\frac{d^2k_\parallel}{(2\pi)^2} \, 
\frac{e^{i \omega (x_0-x'_0) + i k_\parallel \cdot
(x_\parallel-x'_\parallel)}}{2(|k_\parallel| + \zeta |\omega|)}\;.
\end{equation}

Propagating modes on the plate are then to be found as the singularities in the real
time version of the propagator in momentum space.   
They are then solutions to the equation:
\begin{equation}\label{eq:sing}
i \zeta |\omega| + i \sqrt{-\omega^2 + {\mathbf k}^2_\parallel} = 0 \;,
\end{equation}
Those solutions are: $\omega = \pm  v \;|{\mathbf k}_\parallel|$, where $v \equiv 1/\sqrt{1 + \zeta}$.
They are modes with  $v < c$, for $\zeta>0$. 
Thus, a natural interpretation of $\zeta$, in this context, is that it
represents (in an EM analogy) a plate where $\epsilon > 1$, where $1 + \zeta = \epsilon$. 

A simpler derivation can be obtained by studying the classical equations of motion
for the vacuum field, in the presence of the mirror at $x_3=0$. Fourier
transforming the equation with respect to the time and $x_\parallel$, we
see that the Fourier transformed vacuum field should satisfy:
\begin{equation}
[\partial_3^2 -k_\parallel^2 + \omega^2 - \tilde{\lambda}(\omega)
\delta(x_3)] \,
{\tilde \varphi}(x_3,\omega, k_\parallel) \,=\,0 \;. 
\end{equation}
Thus, nontrivial solutions for ${\tilde \varphi}(x_3,\omega, k_\parallel)$ must be of the form:
\begin{equation}
{\tilde \varphi}(x_3,\omega, k_\parallel) \,=\, f(\omega, k_\parallel) \,
e^{-\sqrt{k_\parallel^2 -\omega^2} |x_3|} \;, 
\end{equation}
which, in order to satisfy the discontinuity imposed by the $\delta$
function, requires:
\begin{equation}
2 \sqrt{k_\parallel^2 -\omega^2} \,=\, \tilde{\lambda}(\omega) \;,
\end{equation} 
which has the same solutions that (\ref{eq:sing}).

Note that there are also solution which are odd in $x_3$, and therefore
they do not see the $\delta$ function. However, since the vacuum field
vanishes on the plane, they do not propagate modes on the mirror.

\end{document}